\documentclass[apjl]{emulateapj}
\usepackage[]{graphicx}
\pdfoutput=1

\newcommand{\kms}{\hbox{${\rm km\;s}^{-1}$}}
\newcommand{\sigsky}{\ensuremath{\sigma_{\rm sky}}}
\newcommand{\mulim}{\ensuremath{\mu_{\rm crit}}}
\newcommand{\Msun}{\ensuremath{M_{\sun}}}
\newcommand{\hi}{H~\textsc{i}}

\slugcomment{To appear in ApJ Letters}

\shorttitle{Outer Disks in Virgo and Field S0 Galaxies}
\shortauthors{Erwin, Guti\'{e}rrez, \& Beckman}

\begin{document}

\title{A Strong Dichotomy in S0 Disk Profiles Between the Virgo Cluster 
and the Field}

\author{Peter Erwin}
\affil{Max-Planck-Institut f\"{u}r extraterrestrische Physik,
Giessenbachstrasse, D-85748 Garching, Germany\\
Universit\"{a}ts-Sternwarte M\"{u}nchen, Scheinerstrasse 1,
D-81679 M\"{u}nchen, Germany}
\email{erwin@mpe.mpg.de}
\and
\author{Leonel Guti\'{e}rrez\altaffilmark{1,2}, John E. Beckman\altaffilmark{2,3}}
\affil{Instituto de Astrof\'{\i}sica de Canarias, C/ Via L\'{a}ctea s/n, 38200 La Laguna, Tenerife, Spain}

\altaffiltext{1}{Universidad Nacional Aut\'onoma de M\'exico, Instituto de Astronom\'ia, Ensenada, B. C. M\'exico}
\altaffiltext{2}{Facultad de F\'{\i}sica, Universidad de La Laguna, Avda. Astrof\'{\i}sico Fco. S\'{a}nchez s/n, 38200, La Laguna, Tenerife, Spain}
\altaffiltext{3}{Consejo Superior de Investigaciones Cient\'ificas, Spain}

\begin{abstract} 

We report evidence for a striking difference between S0 galaxies in
the local field and in the Virgo Cluster. While field S0 galaxies have disks
whose surface-brightness profiles are roughly equally divided between the three
main types (Types I, II, and III: single-exponential, truncated, and
antitruncated), Virgo S0s appear to be entirely lacking in disk truncations.
More specifically, the fraction of truncations in S0 galaxies with $M_{B} < -17$ is
$28^{+7}_{-6}$\% for the field, versus $0^{+4}_{-0}$\% for the Virgo
Cluster galaxies; the difference is significant at the 99.7\% level. The
discrepancy is made up almost entirely by Type~I profiles, which are almost
twice as frequent in the Virgo Cluster as they are in the field.

This suggests that S0 formation may be driven by different processes
in cluster and field environments, and that outer-disk effects can be useful
tests of S0 formation models.

\end{abstract}

\keywords{galaxies: structure --- galaxies: elliptical and 
lenticular, cD --- galaxies: evolution --- galaxies: clusters: general}

\section{Introduction} 

One of the most interesting unresolved questions in the study of galaxy
morphology is the origin of S0 galaxies. These galaxies are common in the denser
parts of clusters \citep{dressler80}, which  suggests a cluster-based formation
mechanism.  The discovery that clusters evolve from spiral-dominated at $z
\gtrsim 0.5$ to S0-dominated at lower redshifts
\citep[e.g.,][]{dressler97,postman05,poggianti08} is evidence that clusters may
in fact be turning spiral galaxies into S0s. Indeed, transformation mechanisms
such as ram-pressure stripping \citep{gunn-gott72} require high orbital
velocities and a relatively dense, hot IGM, as found in clusters.

However, S0 galaxies are also found in \textit{low}-density environments such as
small groups and even the local field
\citep[e.g.,][]{vandenbergh09,wilman-erwin11,calvi11}, which suggests one of two
things: either cluster-specific mechanisms such as ram-pressure stripping cannot
create S0s, or there are actually multiple channels for forming S0s,
some operating in clusters and others in lower-density regions.  \citet{moran07}
found evidence suggesting that both ram-pressure stripping and
strangulation/starvation \citep{larson80} may be operating in massive clusters,
while \citet{wilman-erwin11} argue that a significant fraction of local field
and group S0s probably became S0s as central galaxies within their dark-matter
halos, which implies mechanisms other than interactions with the IGM
\citep[e.g., minor mergers; see][]{bekki98,bournaud05}. If the multiple-channel
scenario is true, then there may be differences in S0 galaxies as a function of
environment, differences which could be used to test S0 formation models.

\begin{figure*}
  \centering
  \includegraphics[width=7.2in]{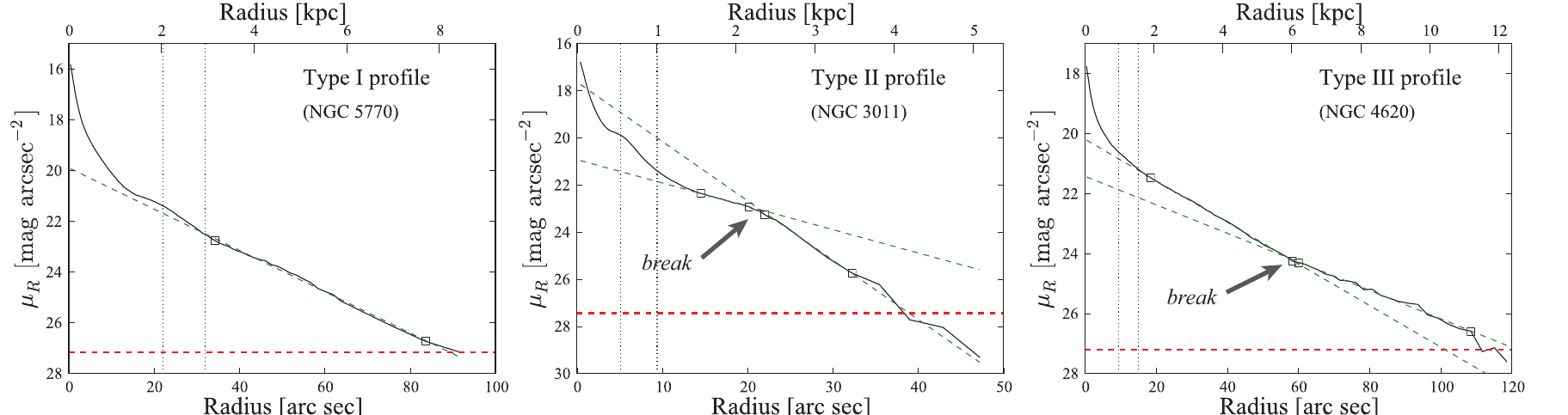}
  
\caption{Examples of the three surface-brightness profile classes.  Type~I
(single-exponential) and Type~II (truncated) profiles are for field S0s;
the Type~III (antitruncated) profile is from the Virgo Cluster S0 
NGC~4620. Diagonal dashed lines (green in online version) are exponential
fits to local regions of the profiles (bounded by small
boxes); vertical dotted lines indicate lower and upper limits to bar size.
Arrows indicate breaks in Type~II and Type~III profiles. Horizontal dashed
lines (red in online version) show the sky-uncertainty limit $\mulim$.
}\label{fig:profiles}

\end{figure*}

One galaxy feature which might show environmental influences quite clearly is
the structure of the outer disk, which should be more vulnerable to effects such
as gas stripping, tidal perturbations, and late-time accretion. Recent work
\citep{erwin05,pt06,erwin08,gutierrez11} has shown that the surface-brightness
profiles of outer disks -- traditionally supposed to be purely exponential, or
else exponential with sharp outer truncations -- fall into three broad classes:
single-exponential (\nocite{freeman70}Freeman [1970] Type~I), truncated (Freeman
Type~II), or antitruncated \citep[Type~III;][]{erwin05}; see
Figure~\ref{fig:profiles} for examples. All three profiles have been observed in
spiral galaxies out to $z \sim 1$
\citep{perez04,trujillo05,azzollini08,bakos11}. The analysis of
\citet{gutierrez11} indicated that local S0 disks are evenly divided between the
three main classes. That study, however, was based on a sample of galaxies 
including both the Virgo Cluster and the local field. Clearly, it would be of
interest to see whether there are differences in the outer disks of S0 galaxies
as a function of environment. This Letter explores that possibility by comparing
the outer disk profiles of S0s in the Virgo Cluster with those of local field
S0s.

\section{Sample Selection and Data} 

\subsection{Virgo and Field Samples}\label{sec:samples}

The Virgo sample was constructed using the Virgo Cluster Catalog
\citep[VCC;][]{vcc}, starting with all galaxies having $B < 16$ \textit{and}
secure membership.\footnote{The VCC classifies galaxies as ``certain members'',
``possible members'', and ``background galaxies''; we only considered
``certain'' members.}  Data from RC3 \citep{rc3} were then obtained for each of
these galaxies, in order to isolate galaxies which were S0 \textit{and}
had axis ratios $\leq 2.0$.  (NGC 4733 is classified as ``E+'' in RC3 but has an
SB0 classification in NED and clearly hosts a strong bar, so we included it in
the sample.)  We then checked each galaxy for distance measurements. This showed
that NGC 4600 had a surface-brightness fluctuation distance of $\sim 7$
Mpc \citep{tonry01}, indicating that is a foreground object; we transferred it
to the field sample. Several other galaxies were rejected because they
overlapped with bright neighbors or very bright stars, had significantly
distorted outer isophotes, were clearly edge-on, or were not actual S0 galaxies.
Finally, we restricted ourselves to galaxies with $M_B < -17$; this limit
approximates the traditional boundaries between dwarf and giant galaxies, and
corresponds to stellar masses $\gtrsim 10^{9.5} \Msun$ (assuming a typical S0 $B
- V$ of $\sim 0.8$ and the color-based mass-to-light ratios of
\nocite{zibetti09}Zibetti et al.\ 2009). The result was a set of 24 Virgo
Cluster S0 galaxies.

The field sample was designed to have the same absolute magnitude limit and a
distance limit of $\sim 30$ Mpc.  The starting point was a combined set of
barred and unbarred galaxies, whose surface-brightness profiles were presented
in \citet{erwin08} and \citet{gutierrez11}, respectively (many of the Virgo S0
galaxies were also included in those studies).  However, those samples were
diameter-limited ($D_{25} \geq 2.0\arcmin$) and therefore tended to exclude more
compact galaxies and galaxies at larger distances. To better match the Virgo
sample, we defined the field sample as follows: all galaxies from HyperLeda with
Hubble types $-4 \leq T \leq 0$, $M_{B} < -17$, declination $\delta >
-10\arcdeg$, and redshifts (corrected for Virgocentric infall) $< 2000$ \kms.
Galaxies which were VCC members were excluded (the only exception being
NGC~4600; see above). As in the Virgo Cluster sample, RC3 measurements were then
used to identify S0 galaxies with axis ratios $\leq 2.0$. After closer
inspection, nine galaxies were rejected for having been misclassified (e.g., NGC
2853 is classified as SB$0^{0}$, but is clearly an early-type spiral),
interacting or overlapping with neighboring galaxies or bright stars, or being
too highly inclined despite the published axis ratio.

The resulting (nominal) field sample has a total of 55 S0 galaxies; however,
suitable images are not available for all.  Since images from the Sloan
Digital Sky Survey \citep[SDSS;][]{york00} have proven more than adequate
for outer-disk profile work \citep[see][]{pt06,erwin08}, we decided
to use Data Release 7 \citep[DR7][]{abazajian09} as our primary image source.
This encompasses the entire Virgo Cluster and much of the nearby northern field;
43 of the field S0 galaxies have DR7 images.  An additional seven S0 galaxies
outside DR7 have already been analyzed using other image sources by
\citet{erwin08} and \citet{gutierrez11}. We include these in our final sample in
order to increase the field-galaxy numbers.  Doing so does, however, introduce a
slight bias in favor of brighter galaxies in the field sample, since the
galaxies in the Erwin et al./Gutierrez et al. samples were selected to have
$D_{25} \geq 2.0\arcmin$. Therefore, we check all our results by using both the
full field sample (50 S0 galaxies) and a ``DR7-only'' subsample, restricted to
the 43 field S0 galaxies covered by DR7.

We use the term ``field'' somewhat loosely, since we did not attempt to
distinguish between groups and genuinely isolated galaxies. However,
we \textit{do} believe that there is a clear difference between the Virgo
Cluster and field samples: the latter does not include any truly high-mass
groups comparable to Virgo. In the Nearby Optical Galaxy Groups (NOGG) catalogs
\citep{giurucin00}, the largest group within 2500 \kms{} in redshift and
with $\delta > -10\arcdeg$ \textit{other} than the Virgo Cluster is the Ursa
Major ``Cluster'' \citep{tully96}, which has three of our field S0s. This
spiral-dominated group has a velocity dispersion of only 148 \kms, compared
with 715 \kms{} for the Virgo Cluster \citep{tully87}. Twelve field S0s have no
group assignment in the NOGG catalogs at all and are therefore plausible
isolated galaxies. The field sample is thus a mixture of low-mass groups and
isolated galaxies, providing a strong contrast with the Virgo Cluster
sample. The two samples do not show any significant differences in
absolute magnitude (Kolmogorov-Smirnov $P = 0.79$).

\section{Analysis and Disk-Profile Generation}\label{sec:analyze}

Details of the process for extracting azimuthally averaged surface-brightness
profiles of outer disks are presented in \citet{erwin08}. Here, we briefly
summarize the basic approach and discuss new automated techniques used to
analyze SDSS images of the 28 galaxies not previously classified by Erwin et
al.\ or \citet{gutierrez11}. Images for each galaxy were retrieved from the SDSS
archive; if the galaxy was near the top or bottom of the image, the adjacent
field was retrieved and the fields merged to form a larger image (adjacent
fields in the vertical direction are from a single drift-scan observing run).

Careful sky subtraction is essential, as is careful masking of bright stars and
other galaxies near the target galaxy. We started by running SExtractor
\citep{bertin96} on the $r$-band images. The resulting catalog was processed
into an SAOimage DS9 region file (converting stars to circles and galaxies to
ellipses, with sizes being multiples of the SExtractor isophotal
area)\footnote{Based partly on code kindly provided by Michael Pohlen.}. These
regions were then displayed on the image and edited (adding additional masking
if necessary and removing regions corresponding to the target galaxy). A mask
image was then generated from the region file, for use in conjunction with the
ellipse-fitting (below).  The mask file was \textit{also} used for sky
subtraction: median pixel values were determined for 100--150 $10 \times
10$-pixel regions, with locations chosen randomly such that they fell outside
masked regions \textit{and} were not within $2.5 \, R_{25}$ of the galaxy center
($R_{25}$ = half of the $D_{25}$ diameter). The final sky value was the mean of
these median measurements, with uncertainty \sigsky{} computed by bootstrap
resampling. We used \sigsky{} to compute a limiting surface brightness $\mulim =
4.94 \sigsky$; this is the level at which a 1-\sigsky{} error in the sky
subtraction would produce a 0.2 mag arcsec$^{-2}$ shift in the
surface-brightness profile \citep[see][]{pt06,erwin08}. Median-smoothed images
were inspected to identify any residual large-scale gradients (typically
vertical, due to changes in sky brightness during the SDSS drift scan); if
present, these were removed with the \textsc{iraf} \textsc{imsurfit} task.

After sky subtraction, the galaxy disk orientation was determined by fitting
ellipses to the $r$-band isophotes, using the \textsc{iraf} task
\textsc{ellipse}. Ellipses corresponding to the outer disk (i.e., outside any
bars and prominent rings) were used to determine the overall galaxy orientation,
under the assumption that the outer disk is approximately circular. We then
re-ran \textsc{ellipse} with ellipse shape and orientation fixed to that of the
outer disk. This generates surface-brightness profiles corresponding to circular
averaging if the galaxy were face-on.  (This approximation fails at small radii
if there is a prominent, rounder bulge, but this does not affect the outer
profile.) A key advantage of this approach is that it allows measuring the
profile to much fainter levels than is possible when the program must
simultaneously fit variable ellipses to the isophotes; typically, we can trace
the profile to at least twice $R_{25}$.

Finally, the surface-brightness profiles were classified into the three
basic types (Figure~\ref{fig:profiles}), with additional subtyping as
described in \citet{pt06} and \citet{erwin08}. Classifications and
parameters for galaxies not in \citet{erwin08} or \citet{gutierrez11} are
presented in Table~\ref{tab:results}.

Since the influence of environmental effects are expected to be stronger in the
outer disk, it makes sense to concentrate on differences in profiles at larger
radii. \citet{erwin08} and \citet{pt06} noted that a small fraction of Type~II
profiles had breaks at or interior to the bar radius (``Type~II.i'' profiles).  
It seems likely that these profiles are not produced by the same mechanisms that
produce truncations at much larger radii; in fact, they appear to be produced
spontaneously in isolated disks due to bar formation \citep{erwin12}. Thus, in
our statistical analyses we do not consider II.i profiles to be true
``truncations'', and group them with the Type~I or III profiles (depending on
the profile shape outside the bar). Because the distinction between II.i and
II.o profiles is only possible for barred galaxies, we also look at the
distribution of profile types for barred galaxies separately. For the S0
galaxies not previously studied in \citet{erwin08} or \citet{gutierrez11}, we
carefully analyzed red (and, if available, near-IR) images for the presence of
bars, and measured any such bars that were found; these measurements will be
presented elsewhere. The absolute numbers of Type~II.i profiles are quite small:
two in the Virgo Cluster S0s and none in the field.

\section{Comparisons Between Virgo and Field Galaxies}\label{sec:compare}

\begin{figure*}
  \centering
  \includegraphics[width=7.0in]{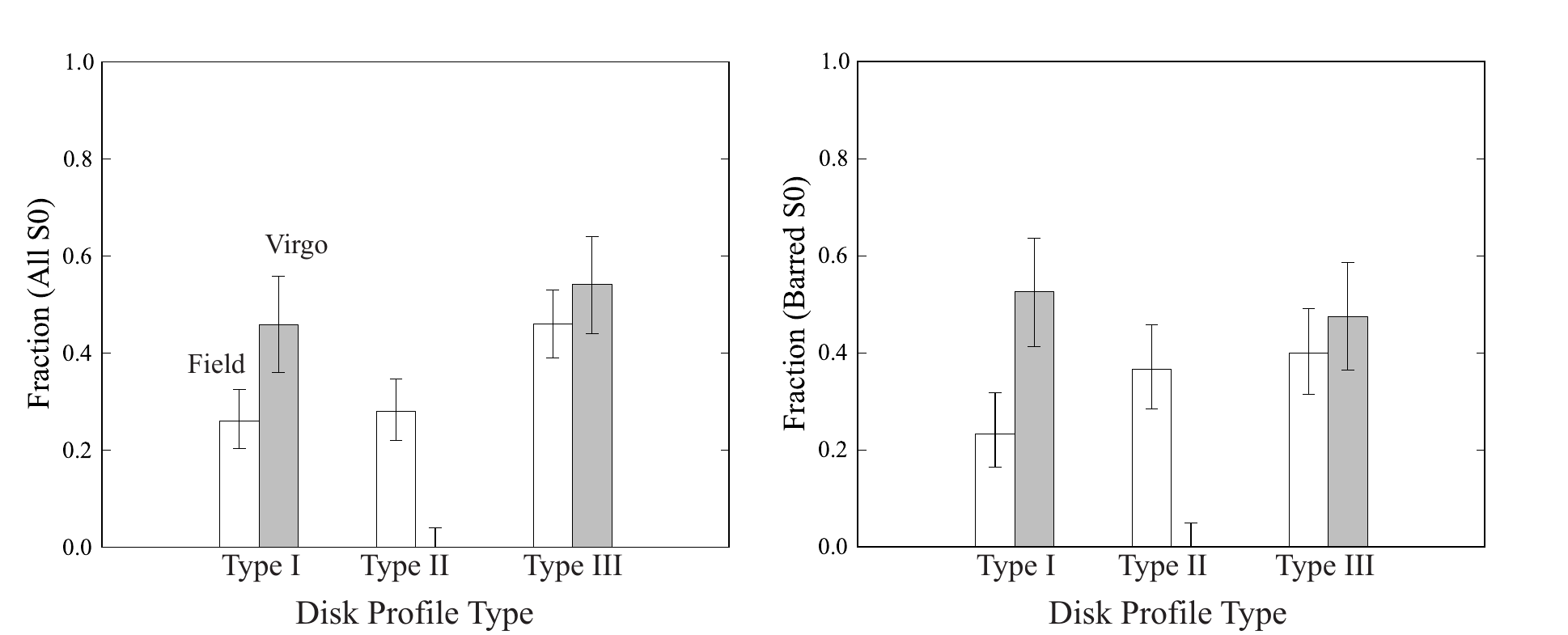}
  \caption{Histograms of outer-disk profile frequencies for S0 galaxies
  in the local field (white bars) and the Virgo Cluster (gray bars). Left:
  all S0 galaxies. Right: barred S0s only. Error bars
  show 68\% confidence intervals \citep{wilson27}.}
 \label{fig:hist}
\end{figure*}

The distribution of profile types is summarized in Figure~\ref{fig:hist}, both
for the complete sample (left panel) and for just the barred galaxies
(right panel). Two features are immediately apparent. First, there is a
\textit{complete absence of truncations in Virgo Cluster S0s}, in contrast to
the relatively high frequency of truncations in field S0s:
$28.0^{+6.7}_{-5.9}$\% of all field S0s ($36.7^{+9.1}_{-8.2}$\% of barred field S0s).
For the DR7-only subsample (see Section~\ref{sec:samples}), the frequencies in
the field are $29.5^{+7.3}_{-6.4}$\% for all field S0s and
$38.5^{+9.8}_{-8.9}$\% for the barred S0s. Second, Type~I
profiles are significantly more common in the Virgo Cluster ($45.8^{+10.1}_{-9.5}$\%)
than in the field ($26.0^{+6.6}_{-5.7}$\%, or $23.3^{+8.5}_{-6.8}$\% for
just the barred galaxies; $22.7^{+6.9}_{-5.7}$\% and $19.2^{+8.8}_{-6.5}$\%,
respectively, for the DR7-only subsample). Type~III profiles, on the other hand,
are equally common in both environments.

The difference in profile frequencies seems strong; is it statistically
significant? Rather than focus on, e.g., just the difference in truncation
frequency (which amounts to selecting out the strongest deviation after the
fact), we ask whether the distribution of all three profile types is similar for
Virgo and field galaxies. For this, we can use the extended version of
Fisher's Exact Test. For the complete Virgo and Field samples, the $3 \times 2$
Fisher's Exact Test\footnote{We use the implementation in the R statistical
package (http://www.r-project.org/)} gives $P = 0.00623$ for the null hypothesis
that both field and Virgo distributions come from the same parent population,
and $P = 0.00366$ when only the barred galaxies are considered. Restricted to
the DR7-only subsample, the signficance becomes even stronger: $P = 0.00318$ for
all S0s and $P = 0.00264$ for barred S0s. (When the smaller number of
\textit{unbarred} galaxies are considered, the differences are \textit{not}
significant.) So the difference in outer-disk profiles between the Virgo Cluster
and the local field is both large \textit{and} statistically significant.

\begin{figure}
  \centering
  \includegraphics[width=3.7in]{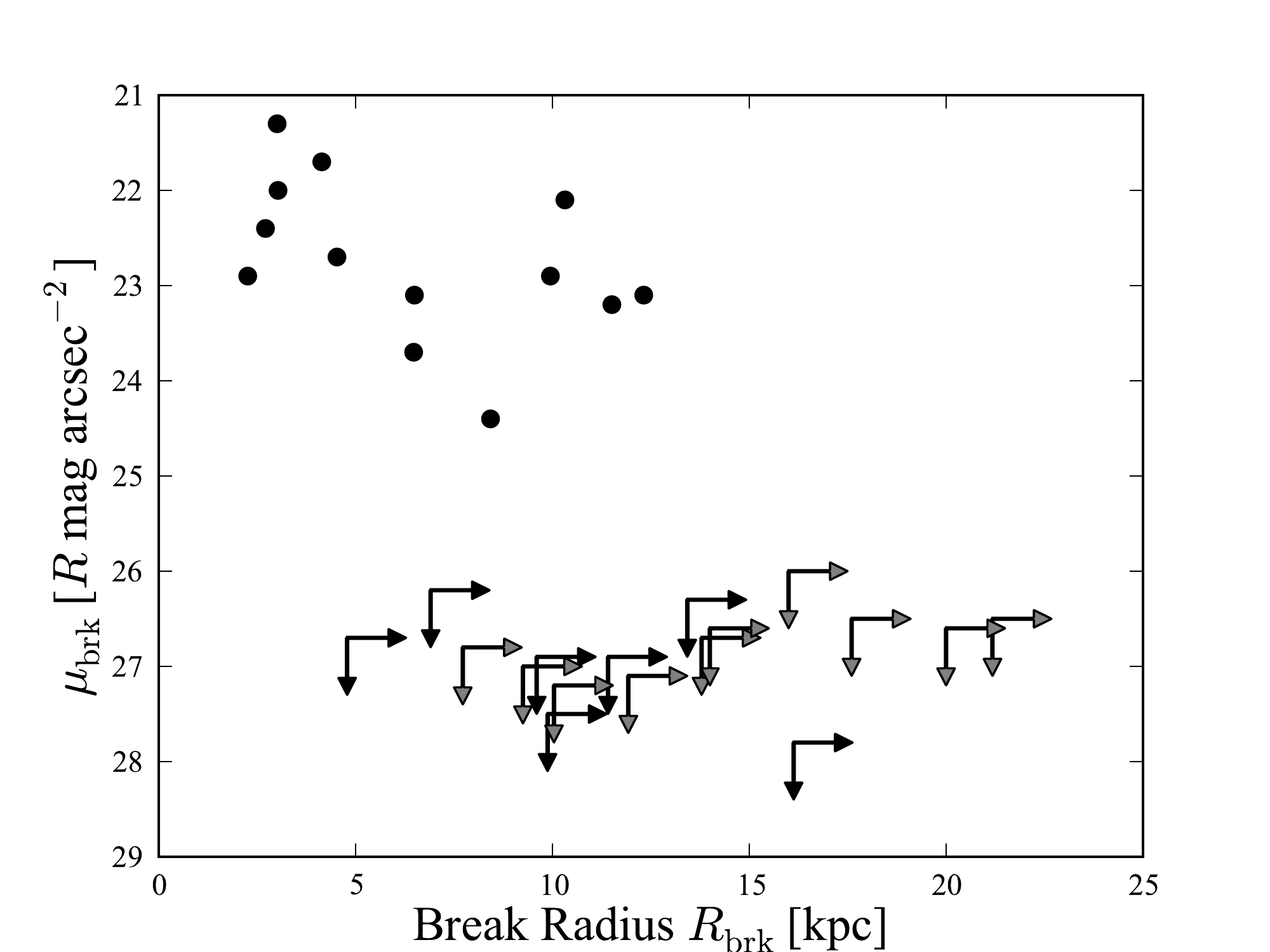}
  
\caption{$R$-band surface brightness at the break ($\mu_{\rm brk}$) versus break
radius $R_{\rm brk}$ for truncations (black circles). Also shown are upper
limits on $\mu_{\rm brk}$ and lower limits on $R_{\rm brk}$ for undetected
truncations in Type~I profiles (black = field, gray = Virgo Cluster).  The clear
gap in $\mu_{\rm brk}$ between the detected truncations and the upper limits
indicates that we are probably \textit{not} missing truncations in Virgo
Cluster galaxies due to surface-brightness limits.}\label{fig:break-radii}

\end{figure}

Could we be missing truncations at larger radii in the Virgo S0s, and
thus mis-classifying them as Type~I?  This seems quite unlikely.
Figure~\ref{fig:break-radii} shows that break radii for field S0s occur
at $R \sim 2$--12 kpc, with surface brightnesses at the break of $\mu_{\rm brk, R}
< 24$ mag arcsec$^{-2}$; the mean $\mu_{\rm brk, R}$ is 22.7.  Also plotted are
limits on any possible breaks for Type~I profiles. The latter are all $>
26$ mag arcsec$^{-2}$, much fainter than any observed breaks.

\section{Discussion}\label{sec:discuss}

We can summarize the difference between outer-disk profiles in the Virgo Cluster
and the field thus: Type~III profiles are (roughly) equally common in Virgo and
in the field; in Virgo, the remaining profiles are \textit{all} Type~I, while in
the field they are half Type~I and half Type~II.  (We note that
\nocite{maltby11}Maltby et al.\ 2011 found no significant environmental differences 
for \textit{spiral} disk-profile frequencies in and around a
multiple-cluster system at $z \sim 0.17$; however, their classification scheme
used a narrow surface-brightness range and is thus rather different
from ours.\footnote{For example, about half of their ``Type~I$_{o}$'' example
profiles show clear Type~II breaks.}) If we focus on the difference between Type~I
and II profiles and assume that it is indeed directly due to environmental
effects, we can posit two possibilities: either something in the cluster (or
proto-cluster) environment transforms Type II profiles into Type~I, or else
something \textit{prevents} a Type~I to Type~II transition which is common in
the field.

The most popular models for disk truncations combine two elements. The
first is a radial dropoff in efficient star formation, either because the gas
density falls below some critical star-formation threshold
\citep{kennicutt89,schaye04,elmegreen06} or because the gas density profile
itself has a sharp break, possibly due to accretion-induced warping
\citep{roskar08a,sanchez-blazquez09,martinezserrano09}. The second element is
the outward scattering of stars from the inner disk, such as produced by
transient spiral arms
\citep{sellwood02,roskar08a,roskar08b,sanchez-blazquez09,martinezserrano09}. The
main problem is that these models tend to approach disk truncation as a
universal phenomenon and do not, in general, explain how and why truncations
might \textit{not} occur.

If we assume that a significant fraction of present-day cluster S0s originally
had truncations, then we can look for something which could erase truncations.
One possibility is ``harassment'' \citep[e.g.,][]{moore96,moore99},
where a galaxy's motion through the (evolving) cluster potential leads to
repeated tidal shocks. The ``high-surface-brightness'' model galaxy in
\citet[][their Fig.~7]{moore99} had its initially single-exponential profile
transformed into a mild antitruncation by this process; this suggests that a
Type~II profile could ``flatten out'' into a Type~I profile \citep[but
see][]{gnedin03}. More detailed simulations involving disks with initially
truncated profiles are needed to see if this is a viable mechanism. (Of course,
this cannot be the \textit{only} way to form Type~I profiles, since they are
also found in the field.)

Alternatively, we can consider cluster-based mechanisms which might prevent
Type~II profiles from forming in the first place.  If, as suggested by
\citet{erwin08} and \citet{erwin12}, truncations in S0s and early-type spirals
are predominantly related to the same sort of bar--Outer Lindblad Resonance
(OLR) interactions which produce outer rings, then one possible scenario might
be the following. The long dynamical times in outer disks suggest -- and
simulations agree -- that large-scale changes there such as outer-ring
formation require several Gyr \citep[see references in][]{buta96}. Since bar-OLR
interactions are strengthened by the presence of signficant gas in the outer
disk (gas, being dynamically cooler than the stellar disk, more readily
absorbs angular momentum from the bar), the \textit{removal} of gas should weaken
OLR effects.  One possible signature of this might be a tendency of
gas-deficient spirals to be less likely to have Type~II profiles and outer
rings.\footnote{The referee pointed out the example of NGC~4245, a strongly
barred field S0/a with a Type~I profile \citep{erwin08} and strong \hi{}
deficiency \citep{boselli09}.} If S0s in the field were able to retain
gas in their outer disk for longer periods, they would be more likely to
show the effects of bar-OLR interactions and develop Type~II profiles. S0
galaxies in Virgo, on the other hand, could have lost their gas -- particularly
in the outer disk -- earlier on due to, e.g., a combination of
ram-pressure stripping and strangulation.

\acknowledgements

We thank Dave Wilman for helpful comments, Michael Pohlen for the initial
version of the catalog-parsing code, and the referee for suggesting the possible
relevance of \hi{} depletion. P.E. was supported by DFG Priority Program 1177.



\begin{deluxetable}{llrrrrrr}
\tablewidth{0pt}
\tablecaption{New Outer Disk Classifications and Measurements\label{tab:results}} 
\tablecolumns{8}
\tablehead{
\colhead{Galaxy} & \colhead{Profile Type} & \colhead{$h_{i}$} & \colhead{$h_{o}$} &
\colhead{$R_{\rm brk}$} & \colhead{$\mu_{0,i}$} & \colhead{$\mu_{0,o}$} &
\colhead{$\mu_{\rm brk}$} \\
 & & ($\arcsec$) & ($\arcsec$) & ($\arcsec$) & & & }

\startdata
\multicolumn{8}{c}{Virgo Cluster} \\
 NGC 4262 &                  III &  14.1 &  24.9 &      83 & 19.36 & 22.13 &  25.8  \\
 NGC 4306 &             III-d(?) &  11.3 &  19.3 &    27.0 & 19.69 & 20.75 &  22.2  \\
 NGC 4377 &                    I &  11.3 & \nodata &   $>$90 & 18.79 & \nodata & $>$ 26.8  \\
 NGC 4379 &                    I &  17.5 & \nodata &  $>$120 & 19.98 & \nodata & $>$ 27.0  \\
 NGC 4468 &             III-s(?) &  17.2 &  39.7 &      63 & 20.48 & 22.74 &  24.3  \\
 NGC 4476 &             III-s(?) &  16.1 &  24.6 &      50 & 19.96 & \nodata & \nodata  \\
 NGC 4479 &                    I &  23.2 & \nodata &  $>$119 & 21.72 & \nodata & $>$ 27.2  \\
 NGC 4483 &                III-d &  12.6 &  20.4 &    55.0 & 19.10 & 20.89 &  23.7  \\
 NGC 4528 &                III-s &   7.4 & \nodata &      29 & 17.60 & \nodata &  21.7  \\
 NGC 4598 &         II.i + III-d &  13.6 & \nodata &    54.0 & 19.76 & \nodata &  24.0  \\
          &                    &  13.6 &  21.7 &    54.0 & 19.76 & 21.38 &  24.0 \\
 NGC 4620 &                III-d &  15.7 &  22.9 &    57.5 & 20.19 & 21.43 &  24.1  \\
 NGC 4733 &                  III &  16.8 &  23.0 &    59.0 & 19.62 & 20.63 &  23.3  \\
 \multicolumn{8}{c}{Field} \\
 NGC 3011 &          II.o-OLR(?) &  11.0 &   4.3 &    21.0 & 20.76 & 17.43 &  22.9  \\
 NGC 3156 &             III-d(?) &  12.0 &  16.9 &      41 & 18.99 & 20.04 &  22.6  \\
 NGC 3266 &             II.o-OLR &  17.4 &  10.3 &    24.9 & 20.41 & 19.36 &  22.0  \\
 NGC 3643 &             III-d(?) &  11.6 &  17.3 &      43 & 19.76 & 22.12 &  24.9  \\
 NGC 3757 &             III-d(?) &  11.6 &  48.8 &    48.5 & 20.40 & 23.87 &  24.7  \\
 NGC 3773 &                    I &  11.9 & \nodata &   $>$70 & 20.20 & \nodata &  26.7  \\
 NGC 3870 &                III-d &   8.3 &  17.8 &      38 & 19.40 & 22.30 &  24.0  \\
 NGC 4221 &             II.o-OLR &  38.3 &  15.1 &      82 & 18.46 & 22.02 &  24.4  \\
 NGC 4391 &             II-CT(?) &  12.3 &   6.4 &      24 & 20.14 & 18.21 &  22.4  \\
 NGC 5507 &             III-s(?) &  10.9 &  35.4 &      65 & 18.70 & \nodata & \nodata  \\
 NGC 5631 &             III-s(?) &  18.3 & \nodata &      45 & 19.62 & \nodata & \nodata  \\
 NGC 5770 &                    I &  13.3 & \nodata &   $>$77 & 19.90 & \nodata & $>$ 26.2  \\
 NGC 5839 &                    I &  14.0 & \nodata &   $>$90 & 19.91 & \nodata & $>$ 26.9  \\
   IC 745 &                  III &   5.3 &   7.4 &      28 & 18.81 & 20.42 &  24.4  \\
  IC 2450 &                III-d &  10.3 &  12.3 &      34 & 19.66 & 20.15 &  23.3  \\
 UGC 5745 &             III-s(?) &   8.4 &  14.2 &      47 & 19.03 & \nodata & \nodata  \\
 UGC 9519 &                  III &   5.6 &   8.7 &      26 & 19.02 & 20.82 &  24.1  \\
\enddata 

\tablecomments{New classifications and measurements of $R$-band outer-disk
parameters for S0 galaxies \textit{not} in \citet{erwin08} or
\citet{gutierrez11}.  For each galaxy, we list profile type, exponential scale
lengths of inner and outer parts of the profile, break radius $R_{\rm brk}$ for Type~II and
III profiles, central surface brightnesses of fitted exponentials, and surface
brightness at the break radius $\mu_{\rm brk}$ (see Erwin et al.\ for details). 
Surface brightnesses are observed values (uncorrected for Galactic extinction,
inclination, or redshift). Type~I profiles by definition have no ``outer'' part
and only upper limits for $R_{\rm brk}$ and $\mu_{\rm brk}$.}

\end{deluxetable}

\end{document}